\documentclass[usenatbib]{mn2e}

\usepackage{times}
\usepackage{graphicx}

\title[Reflection Studies of XTE~J1118$+$480]{Extremely Weak
Reflection Features in the X-ray Spectrum of XTE~J1118$+$480:
Possible Evidence for X-ray-Emitting Jets?}

\author[J. M. Miller et al.]{J.~M.~Miller$^1$, D.~R.~Ballantyne$^2$, A.~C.~Fabian$^2$, and W.~H.~G.~Lewin$^1$\\ 
	$^1$Center for Space Research and Department of Physics,
	Massachusetts Institute of Technology, Cambridge, MA
	02139-4037, USA;\\ 
	jmm@space.mit.edu\\
	$^2$Institute of Astronomy, University of Cambridge, Madingley
	Road, Cambridge CB3 OHA\\}

\begin{document}
\maketitle

\begin{abstract}
We have simultaneously fit \textit{Chandra} and \textit{RXTE} spectra
of the Galactic black hole XTE~J1118$+$480 with three models for X-ray
reflection.  We explored a range of accretion disc ionizations
(log($\xi$)$=$1--4; $\xi=L_{X}/nR^{2}$) and iron abundances
(0.10--1.00).  Our fits with the constant density ionized disc models
of Ross \& Fabian indicate that $\leq$0.5 per cent (90 per cent
confidence upper-limit) of the observed flux is reflected.  Fits with
the ``pexrav'' of model Magdziarz \& Zdziarski indicate that the
two-dimensional solid angle ($\Omega/2\pi$) subtended by the disc
relative to a central source of incident hard X-rays is
$0.01^{+0.06}_{-0.01}$.  A combination of the high inclination
($i=81$ degrees), Comptonization, and bulk velocities may
each contribute to the low reflection fractions we have measured.
The results are also consistent with extended jets being the source
of the hard X-ray flux, as the disc would then represent a small solid
angle as seen from the emission region.
\end{abstract}


\section{Introduction}
Reflection geometries are well-established in AGNs, and the same basic
geometry -- the irradiation of an accretion disc by hard X-rays -- is
very likely at work in Galactic black holes (see, e.g., Gierlinski et
al. 1997; Zdziarksi et al. 1998; Miller et al. 2001).  The shape of
the reflected component in a given spectrum depends on several
factors, including the height and distribution of hard X-ray sources
above the accretion disc, the temperature and ionization of the disc,
and how close to the black hole the reflection occurs.  Thus, the
reflected component of a given spectrum may be used to constrain the
accretion flow geometry.

XTE J1118+480 is particularly well-suited to reflection studies.  In
the bright X-ray states of most Galactic black holes (GBHs), the
accretion disc is an X-ray object.  This soft spectral component
complicates fits with reflection models.  In less luminous states, the
accretion disc temperature is often below the soft X-ray band.  Due to
the extremely high latitude of XTE~J1118$+$480, the disc temperature
could be measured via UV spectroscopy.  At kT = 24 eV (McClintock et
al. 2001a), the inner disc temperature in XTE~J1118$+$480 is similar
to, or lower than, the inner disc temperatures measured in many AGN.
In any reflection analysis, the inclination of the system and the mass
of the black hole are important parameters, and both have been
measured in XTE J1118+480 ($i=81\pm2$~degrees, Wagner et
al. 2001; $M_{BH} > 6.0~M_{\odot}$, McClintock et al. 2001b), allowing
us to more tightly constrain other model parameters.

Two separate new models have been proposed to describe
XTE~J1118$+$480, which should produce different reflection signatures.
Markoff, Falcke, and Fender (2001) have described a model in which
synchrotron emission from the jet in this system (Fender et al. 2001)
can account for the observed radio and X-ray spectra.  In this model,
a significant fraction of the power-law X-ray flux (perhaps all such
flux) may be due to the jet.  In such a geometry, one would expect
very little of the flux to reflect from the accretion disc, and into
the line of sight of the observer.  Noting the similarity of optical
and X-ray variability in this system, Merloni, Di Matteo, and Fabian
(2000) have described a model wherein magnetic loops from the corona
may generate hard X-rays and illuminate the accretion disc, creating
the observed X-ray and optical QPOs.  Although the inclination of
XTE~J1118$+$480 is high, if this process is at work, a fraction of the
incident hard X-rays from the magnetic loops might scatter into the
observing line of sight.

McClintock et al.\ (2001a) and Esin et al.\ (2001) describe the
accretion flow geometry in terms of an advection-dominated accretion
flow (ADAF).  In this picture, the central accretion region is a hot,
quasi-spherical, radiatively-inefficient volume, and the inner
accretion disc edge is truncated at a radius far from the black hole.
A central ADAF may be expected to diminish the strength of any
reflection features (e.g. Gierlinski et al.\ 1997).  In contrast,
Frontera et al.\ (2001) have described the accretion geometry in terms
of a hot accretion disc.  This work considers reflection explicitly
and finds that the angle ($\Omega/2\pi$) subtended by the disc ranges
between 0.11--0.18 for a variety of models.

For the purpose of constraining the accretion flow geometry in
XTE~J1118$+$480, we have fit a number of reflection models, covering a
large parameter space.  \textit{Chandra} and \textit{RXTE} spectra
were fit simultaneously, covering the 0.4--100~keV band.  In the
following sections, we detail our analysis and results.  We also
comment on recent models for XTE~J1118$+$480.

\section{Observation and Data Reduction}
We analyzed the \textit{Chandra} LETG/ACIS-S spectrum reported on by
McClintock et al. (2001a), reduced via the methods detailed in that
work.  \textit{Chandra} observed XTE~J1118$+$480 on 18 April 2000,
between 18:16--02:16 (UT), for an integrated exposure of 27.2~ksec.
We consider the 0.4--5.0~keV range with 5 per cent systematic errors.
Below 0.4~keV and above 5.0~keV, the data are more noisy and may
require larger systematic errors (see the discussion in McClintock et
al. 2001a), so we have restricted our fits to the 0.4--5.0~keV range.

We have summed 18 observations made with \textit{RXTE} between 13
April and 15 May, 2000, for a total of 46~ksec of PCA data and 18~ksec
of HEXTE data.  This summation is not likely to distort any spectral
features due to the extremely steady nature of XTE~J1118$+$480: fits
to individual spectra indicate that the power-law index is constant
at the 3$\sigma$ level of confidence across this time range
(M. P. Muno, priv. comm.).

The PCA data was reduced using FTOOLS version 5.  We used all
available proportional counter units (PCUs; generally, PCUs 0, 2, and
3) in each observation, and selected only the top layers of the
available PCUs.  Response matrices were made using version 2.43 of
``pcarsp.''  Background spectra were made using the 2000 January 31
``blank-sky'' model for gain epoch 4, and version 2.1e of
``pcabackest.''  Background spectra were subtracted using ``mathpha.''
We consider the 2.8--25~keV energy range, as residuals below 2.8~keV
are relatively large and the PCA spectrum becomes background-dominated
above 25~keV.  Spectra from individual observations are summed using
``mathpha.''  Uncertainties in the PCA response may be as large as 1
per cent (Jahoda 2000), so we have added 1 per cent systematic errors
to the statistical errors before fitting the summed PCA spectrum.

The HEXTE data were also reduced using FTOOLS version 5.  The standard
response matrices were used.  The background was estimated using
``hxtback,'' and source spectra were summed using ``mathpha.''  No
systematic errors were added to this data.  Although data in the
15--20~keV range are likely free of any systematic effects, we excluded
this range to avoid any possible response problems at the edge of the
detector energy range.  Even though HEXTE is sensitive up to 250~keV, we
fit only up to 100~keV.  We are interested in comparing reflection
models as consistently as possible, and the version of the ``constant
density ionized disc'' reflection model (Ross \& Fabian, 1993; Ross,
Fabian, \& Young 1999) which calculates a reflection fraction is only
valid up to 100~keV.  The high energy signature of reflection -- the
Compton-backscattering excess -- peaks near $\sim$30~keV (George \&
Fabian 1991), so an upper limit of 100~keV is high enough to allow for
robust analysis.

\section{Analysis and Results}
\textit{Chandra} LETG/ACIS-S, \textit{RXTE} PCA, and \textit{RXTE}
HEXTE-A and HEXTE-B spectra are fit simultaneously, with an overall
normalising constant allowed to float between each.  The fitting range
discussed above (0.4--100~keV) has a total of 1248 energy bins.  For
all fits, we fixed the neutral hydrogen column density to the value
selected by McClintock et al. (2001a): N$_{H}=1.3\times10^{20}~{\rm
cm}^{-2}$.  We note a ``notch'' in the LETG/ACIS-S spectrum near
2~keV.  As this feature was not seen in a \textit{BeppoSAX}
observation of XTE~J1118$+$480 (Frontera et al. 2001), we consider
this feature to be an instrumental artifact.  This assessment is
bolstered by the fact that a similar ``notch'' is seen in other
\textit{Chandra} observations of bright continuum sources (e.g., Patel
et al. 2001, Miller et al. 2002, Juett et al. 2002).  We modeled this
feature with an inverse edge at 2~keV (``$\tau$''$~=~-0.15$).  With
this addition, the LETG/ACIS-S spectral index becomes consistent with
that of the other instruments.
  
Some of the fits we describe below include the effects of relativistic
blurring, which is expected for regimes close to a black hole.  In
XSPEC, it is possible to convolve a fitting model with a Gaussian or
Lorentzian profile.  Instead of convolving with these profiles, we
convolve with a profile appropriate for a general relativistic regime
(the ``diskline'' model within XSPEC).  Where we have blurred the
spectrum, we do so for $6~{\rm R}_{g} \leq {\rm R}_{in} \leq
10^{5}{\rm R}_{g}, {\rm R}_{g}=GM_{BH}/c^{2}$.

The ``pexrav'' and ``pexriv'' models do not include an Fe~K$_{\alpha}$
line, and one must be added explicitly.  The possibility of an iron
line is discussed in McClintock et al. (2001a).  Although we do not
consider the 5--7~keV range in the \textit{Chandra} spectrum due to
larger overall effective area calibration uncertainties in that band,
it is fine for examining the possibility of narrow emission and/or
absorption features.  We therefore adopt the same Fe~K$_{\alpha}$
upper-limit as McClintock et al. (2001a), and constrain the equivalent
width to be less than 24~eV.  We fix the width at zero as any 1--2 bin
features in the LETG are unresolved at the resolution of the PCA.  The
line center is fixed at 6.4~keV.  We also consider the possibility of
an Fe under-abundance intrinsic to XTE~J1118$+$480 in fitting
``pexrav'' and ``pexriv'' to explore the possible effects on the
reflection fraction.  It should be noted, however, that our fits to
the \textit{Chandra}/LETG data in the region of the Fe~L$_{III}$ edge
do not indicate an under-abundance relative to solar values.

Fit parameters for each model which are not discussed below, are
detailed in Table~2.  Unless otherwise noted, all errors quoted in
this paper are the difference between the best-fit value of a given
parameter, and its value at the 90 per cent confidence limits.

\subsection{Fits with a simple power-law}
We first consider a simple power-law model for the spectrum (see
Figure 1).  This simple model is an excellent fit, yielding a very
good reduced fit statistic ($\chi^{2}/\nu=0.982, \nu=1243$).  We
obtain a power-law index of $\Gamma=1.777\pm0.004$, which is
consistent with the values reported by McClintock et al. (2001) for
the \textit{Chandra} LETG/ACIS-S and combined \textit{RXTE} spectra
($\Gamma=1.77\pm0.4$, and $\Gamma=1.782\pm0.005$, respectively).
Frontera et al. (2001) measure a slightly harder power-law index with
\textit{BeppoSAX} ($\Gamma=1.722^{+0.003}_{-0.005}$).  Whereas we fit
an energy range (0.4--100~keV) which does not require a blackbody
component or a cut-off in the power-law slope at high energy, the
\textit{Beppo-SAX} results cover the 0.1--200~keV range, and include a
blackbody component and a cut-off.  Even without these considerations,
it is striking how well these spectral characterisations agree.

\subsection{Fits with ``Pexrav'' and ``Pexriv''}
We next made fits with ``pexrav'' (Magdziarz \& Zdziarski 1995).  This
model describes the reflection of hard, power-law X-rays from a cool,
neutral accretion disc.  The parameters of this model include: the
index of the incident power-law flux, the energy cut-off of the
power-law, the reflection fraction ($0\rightarrow2\pi$, normalised to
1, corresponding to an isotropic source above the disc), the source
redshift, the abundance of elements heavier than He, the abundance of
Fe, the cosine of the inclination angle, and component normalisation
(photon flux at 1~keV).  In our fits, the power-law index and cut-off,
the reflection fraction, and flux were allowed to float.  We fixed the
redshift at zero, the lower abundances at solar, the Fe abundance at
0.10 and 1.00 relative to solar (see below), and the cosine of the
inclination angle to 0.156 ($i=81$~degrees; Wagner et
al. 2001, McClintock et al. 2001b).  We considered this model with and
without incorporating the effects of relativistic blurring on the
overall spectrum.

With the iron abundance set to 1.00 relative to solar abundances, fits
with pexrav return a best-fit reflection fraction of
$f=0.01^{+0.06}_{-0.01}$.  Relativistic blurring does not affect this
result.  However, as XTE~J1118$+$480 is a halo object, it is possible
that it has a significant Fe under-abundance.  We therefore also made
fits with the Fe abundance fixed at 0.10 relative to solar abundance.
Again, the best-fit values for $f$, with blurring and without, were
consistent with zero.  The upper limits with this lower abundance are
$f\leq0.04$ for both blurred and unblurred models.

We conclude that the upper-limits we have obtained with this model
may be imprecise and dominated by the fact that the best-fit
reflection fractions are consistent with zero.  The most reliable
upper-limit on the reflection fraction is likely the largest,
$f\leq0.04$.  With these low reflection fractions, ``pexrav'' is a good
fit to the data (e.g., without blurring and with Fe at solar
abundance, $\chi^{2}/\nu=0.960, \nu=1241$).  

``Pexriv'' (Magdziarz \& Zdziarski 1995) is a cousin to ``pexrav,''
meant to cover regimes in which the accretion disc is ionized.  An
ionized accretion disc might be expected when the inner edge extends
close to the black hole, and is therefore more strongly irradiated by
incident hard X-rays.  For this reason, we only consider this model
with the effect of relativistic blurring.  The parameters of this model
in addition to those for ``pexrav'' include the disc temperature, and
the disc ionization parameter $\xi$.  We fix the disc temperature to
24~eV, as per McClintock et al. (2001a), and examine the extreme
ionization cases: $\xi=0.0,5000$ (5000 is the upper-limit for this
model), and Fe abundances of 0.10 and 1.00 relative to solar.

For solar iron abundances and $\xi=0$, $f=0.01^{+0.06}_{-0.01}$ --
this is exactly what we obtain with ``pexrav'' for the same values.
For all other combinations of Fe abundance and $\xi$, the upper-limit
on the reflection fraction is $f\leq0.001$.  In all cases, ``pexriv''
was an acceptable fit (even for $\xi=5000$); for the $\xi=0$ and Fe
abundance at full solar value, $\chi^{2}/\nu=0.966, \nu=1240$.

We note that a very small error may be introduced by fitting
``pexrav'' and ``pexriv'' with a gaussian iron line model for which
the maximum equivalent width is constrained to be less than 24~eV.
Based on George \& Fabian (1991), in the limit where all of the
observed spectrum is reflected, the Fe~K$_{\alpha}$ line equivalent
width expected at an inclination of $81$~degrees is approximately
40~eV.  However, as we measure only a very small fractional reflection
with ``pexrav'' and ``pexriv,'' we regard any error introduced by
constraining the Fe~K$_{\alpha}$ line EW$\leq$24~eV to be negligible.

\subsection{Fits with the ``Constant Density Ionized Disc Model''}
This model is designed to handle a large range of disc ionizations
(log($\xi$)$=1\rightarrow6$), assuming a constant density accretion
disc and a power-law irradiating spectrum.  The parameters for this
model include the incident power-law flux, the disc ionization, the
source redshift, the reflection fraction, and the component
normalisation (related to photon flux at the detector, but
\textit{not} diluted by 1/d$^{2}$, and therefore on the order of
$10^{-25}$~ph/cm$^{2}$/s).  In our fits, the redshift was fixed at
zero, and all other parameters were allowed to float simultaneously.
We considered this model with the effects of relativistic blurring.

Extremely tight constraints are obtained with this model.  A very low
ionization is indicated: log($\xi$)$=1.056\pm0.002$.  As with the
other models, the best-fit reflection fraction is zero; an upper limit
of $R\leq0.002$ is obtained.  Here, $R$ is the fraction of the total
flux that is reflected (not an angle, like $f$).  The fit achieved
with this model is also good ($\chi^{2}/\nu=1.034, \nu=1241$).  We
tested the robustness of the low values for the ionization parameter
and reflection fraction by fixing these parameters at higher values;
this resulted in large changes in the fit statistic
($\Delta\chi^{2}/\nu\simeq1-3$), indicating that a low ionization and
reflection fraction are strongly preferred with this model.  In Figure
2, we plot the constant-density ionized disk model with $R=0.002,
0.1$, and 0.3 to illustrate the extremely weak nature of the
reflection component we have measured in XTE~J1118$+$480.

The $R$ value measured by this model is an angle-averaged result. By
inspection of the reflection components calculated in George \& Fabian
(1991), it is clear that at 81~degrees the strength of the reflected
component is actually 2-3 times higher than we measure.  Based on this
comparison, the most appropriate upper-limit is slightly higher:
$R\leq0.005$.  

\section{Discussion}
Each of the fits to XTE~J1118$+$480 with ``pexrav'' or ``pexriv''
wherein the strength of the reflection component is measured prefer
$f=0.01^{+0.06}_{-0.01}$ -- consistent with zero.  These measurements
are an order of magnitude lower than those found by Frontera et
al. (2001) in fits to \textit{BeppoSAX} spectra.  Fitting these models
on a range extending to 200~keV for a more direct comparison to the
\textit{BeppoSAX} results only minimally increases the upper limit on
the reflection fraction: $f=0.01^{+0.07}_{-0.01}$.  ``Pexriv'' is not
able to simultaneously constrain the ionization parameter and
reflection fraction, and so we regard results from this model
cautiously.  Indeed, based on the very low inner accretion disk
temperatures measured by McClintock et al. (2001a) and Frontera et
al. (2001) -- $k{\rm T}\simeq24$~eV and $k{\rm T}=32-52$~eV,
respectively -- we expect a largely neutral accretion disk and
therefore ``pexrav'' may be a more appropriate model.  Fits made with
the CDID model were able to constrain the parameters simultaneously,
and indicate very little (if any) reflection component from a neutral
accretion disk ($\xi=11.3\pm0.1$, $R<0.005$), in good agreement with
expectations based on the accretion disk temperature.

A flaring geometry for XTE~J1118$+$480 like that described by Merloni
et al. (2000) is partially supported by our results.  Beloborodov
(1999) examined the effects of irradiating sources with large bulk
velocity on the strength of X-ray reflection.  Such a scenario might
arise if a magnetically-active corona above an accretion disk is
producing flares, and the flares are pushed away from the disk (e.g.,
by the radiation pressure from reprocessed radiation).  For the
measured inclination of XTE~J1118$+$480 ($i=81$~degrees),
there is no bulk velocity ($\beta=0-1, \beta=v/c$) which can explain a
reflection fraction as small as the one we measure.  However, it is
possible that flaring of the kind described by Merloni et al. (2000)
at large scale heights, especially if the flares have a bulk velocity,
might produce a very small reflection fraction.

The effects of an inner ADAF geometry might also diminish the strength
of reflected features.  McClintock et al.\ (2001a) and Esin et al.\
(2001) find ADAF/disc transition radii of $R_{tr}\geq35~R_{Schw.}$ and
$R_{tr}\geq{55}~R_{Schw.}$, respectively.  The degree to which
reflection is diminished due to an inner ADAF can be addressed via
fits to observations of Cygnus~X-1.  The model proposed by Esin et
al.\ (1997) suggests that as the mass accretion rate falls between the
``high/soft'' state and ``low/hard'' states of black hole X-ray
binaries, the inner disc may recede and be replaced by an inner ADAF.
Esin et al.\ (1998) fit an ADAF model to these states in Cygnus~X-1;
in the high/soft state $R_{tr}\sim3.5~R_{Schw.}$, and in the low/hard
state $R_{tr}\geq20~R_{Schw.}$, with a best-fit of
$R_{tr}\sim100~R_{Schw.}$.  Gierlinski et al.\ (1999) and Gierlinski
et al.\ (1997) fit reflection models to the high/soft and low/hard
states, respectively.  The angle subtended by the disc ($\Omega/2\pi$)
reduces from $\sim0.6$ to $\sim0.3$ between these states.  Thus,
although an inner ADAF region of a size comparable to that reported in
XTE~J1118$+$480 reduces by half the strength of the reflected
components observed in Cygnus~X-1, it does not reduce the reflection
to $zero$.  It therefore seems unlikely that an ADAF description of
the accretion geometry in XTE~J1118$+$480 can account for our
reflection results.

If a hot, optically-thin corona partially covers the inner accretion
disk, then at least a fraction of the reflected spectrum must pass
through the corona along a given line of sight.  Petrucci et
al. (2001) examined the effects of this scenario as a function of
inclination angle, assuming a corona which blankets the accretion
disk.  For high inclinations, the strength of the observed reflection
component may be diminished considerably by the effects of
Comptonization.  If the corona covers the accretion disk, then
Comptonization may at least partially explain the very low reflection
fraction in XTE~J1118$+$480.  Such a geometry is not necessarily
favored for this source, however.  Fits to the \textit{Beppo-SAX}
spectrum with a number of Comptonization models by Frontera et
al. (2001) find that the optical depth ($\tau$) of the corona is
\textit{lower} for a ``slab'' geometry than a spherical geometry.  If
the corona is spherical and central, and of a size similar to the ADAF
geometry suggested by McClintock et al. (2001a) and Esin et
al. (2001), then the effects of Comptonization on the reflection
spectrum would likely be rather small.  If the corona were larger and
more diffuse, then a smaller optical depth might be expected for a
spherical volume than for a slab geometry.

Finally, we consider the possibility that the hard, power-law X-ray
flux observed in XTE~J1118$+$480 is due primarily to synchrotron
self-Comptonization in the jet observed in this source (Markoff,
Falcke, \& Fender 2001).  As seen from such an extended emission
region, the accretion disk would subtend a smaller solid angle
than the geometries we consider above.  Of the explanations for a
small reflection fraction that we consider here, this interpretation
can most easily explain why we have measured reflection fractions
consistent with \textit{zero}.  

If the jets are a source of the hard X-ray flux in XTE~J1118$+$480
as our fits may suggest, a number of interesting questions arise.  Are
the X-ray QPOs seen in this source (0.08~Hz; Revnivtsev, Sunyaev, \&
Borozdin 2000) manifested within the jet?  How can the correlated
X-ray and optical variability be explained?  Why is XTE~J1118$+$480
different than other GBHs, wherein non-zero reflection fractions are
measured (see, e.g., Zycki, Done, \& Smith 1997; Gilfanov, Churazov,
\& Revnivtsev 2000), and the jet does not seem to be the major source
of hard X-ray flux?  The answers to these questions can be partially
given by studies of reflection in other GBHs, but may only be answered
definitively by subsequent outbursts of XTE~J11118$+$480 itself.
 
\section{Acknowledgments}
We wish to thank J. E. McClintock, H. L. Marshall, and M. P. Muno for
generously providing data files, and for helpful discussions.  We
acknowledge Andrea Merloni for carefully reading this work and
offering helpful suggestions.  D.~R.~B. acknowledges financial support
from the Natural Sciences and Engineering Research Council of Canada
and a Commonwealth Fellowship.  W.~H.~G.~L. gratefully acknowledges
support from NASA.  This research has made use of the data and
resources obtained through the HEASARC on-line service, provided by
the NASA--GSFC.

\clearpage

\begin{table}

\footnotesize
\begin{center}
\begin{tabular}{llllllllll}
\multicolumn{2}{l}{Observatory} & \multicolumn{2}{l}{Instrument} & Bandpass & Fitting Range & \multicolumn{3}{l}{Date (UT)} & Time (ksec)\\
\hline
\multicolumn{2}{l}{\textit{Chandra}} & \multicolumn{2}{l}{LETG/ACIS-S} &  0.24--7~keV & 0.4--5~keV & \multicolumn{3}{l}{18 Apr 18:16} & 27\\
\multicolumn{2}{l}{\textit{RXTE}} & \multicolumn{2}{l}{PCA} &  2--60~keV & 2.8--25~keV & \multicolumn{3}{l}{13 Apr -- 15 May} & 46\\
\multicolumn{2}{l}{\textit{RXTE}} & \multicolumn{2}{l}{HEXTE} & 15--250 & 20--100~keV & \multicolumn{3}{l}{13 Apr -- 15 May} & 18\\
\hline
\end{tabular} ~\vspace*{\baselineskip}~\\ \end{center}
\caption{We fit the above data from the April--May 2000 outburst of
XTE~J1118$+$480.  We summed 18 observations with \textit{RXTE}, using
all available proportional counter units (PCUs) in each observation
(generally, PCUs 0, 2, and 3).  The power-law slopes from the individual
\textit{RXTE} observations are consistent at the 3$\sigma$ level of
confidence (M. P. Muno, priv. comm.).  Data from both HEXTE clusters
is used.}
\end{table}

\begin{table}
\footnotesize
\begin{center}
\begin{tabular}{llllllllllll}
\multicolumn{2}{l}{Model} & Blurring~$^{a}$ & Fe~$^{b}$ & $\Gamma~^{c}$ & E$_{c}~^{d}$ & \multicolumn{2}{l}{$\xi~^{f}$} & $f_{refl}~^{g}$ & Norm.~$^{h}$ & $\nu~^{j}$ & $\chi^{2}/\nu$\\
\multicolumn{2}{l}{~~} & ~~ & (fixed) & ~~ & (keV) & \multicolumn{2}{l}{~~} & ~~ &~~ & ~~ & ~~ \\
\hline
\multicolumn{2}{l}{power-law} & No & 1.0 & 1.777(4) & -- & \multicolumn{2}{l}{--} & -- & 0.1823(7) & 1243 & 0.982\\
\hline
\multicolumn{2}{l}{pexrav} & No & 1.0 & 1.767(3) & $900^{+500}_{-300}$ & \multicolumn{2}{l}{--} & $0.01^{+0.06}_{-0.01}$ & 0.182(1) & 1241 & 0.960\\
\multicolumn{2}{l}{pexrav} & Yes & 1.0 & 1.767(5) & $900^{+500}_{-300}$ & \multicolumn{2}{l}{--} & $0.01^{+0.06}_{-0.01}$ & 0.178(1) & 1241 & 0.965\\
\multicolumn{2}{l}{pexrav} & No & 0.1 & 1.764(4) & $1000^{+200}_{-300}$ & \multicolumn{2}{l}{--} & $<$0.04 & 0.182(7) & 1241 & 0.961\\
\multicolumn{2}{l}{pexrav} & Yes & 0.1 & 1.767(4) & $900^{+590}_{-300}$ & \multicolumn{2}{l}{--} & $<$0.04 & 0.178(1) & 1241 & 0.966\\
\hline
\multicolumn{2}{l}{pexriv} & Yes & 1.0 & 1.767(4) & $900^{+500}_{-300}$ & \multicolumn{2}{l}{$0^{\dag}$} & $0.01^{+0.06}_{-0.01}$ & 0.178(3) & 1240 & 0.966\\
\multicolumn{2}{l}{pexriv} & Yes & 1.0 & 1.768(4) & $1100^{+600}_{-300}$ & \multicolumn{2}{l}{$5000^{\dag}$} & $<$0.001 & 0.180(3) & 1240 & 0.987\\
\multicolumn{2}{l}{pexriv} & Yes & 0.1 & 1.767(5) & $900^{+600}_{-300}$ & \multicolumn{2}{l}{$0^{\dag}$} & $<$0.001 & 0.178(3) & 1240 & 0.964\\
\multicolumn{2}{l}{pexriv} & Yes & 0.1 & 1.767(5) & $900^{+600}_{-300}$ & \multicolumn{2}{l}{$5000^{\dag}$} & $<$0.001 & 0.178(3) & 1240 & 0.964\\
\hline
\multicolumn{2}{l}{CDID} & Yes & 1.0 & 1.776(2) & -- & \multicolumn{2}{l}{11.3(1)} & $<$0.002 & 3.33(1)$\times10^{-24}$ & 1241 & 1.035\\
\hline
\end{tabular} ~\vspace*{\baselineskip}~\\ \end{center}
\caption{The results of simultaneously fitting the \textit{Chandra} and \textit{RXTE} with reflection models.  All errors are 90 per cent confidence.
$^{a}$ Blurring is a relativistic convolution; see Section 3.
$^{b}$ The iron abundance relative to solar.
$^{c}$ The irradiating power-law index.
$^{d}$ The high energy cut-off of the irradiating flux.
$^{f}$ The ionization parameter, $\xi=L_{X}/nR^{2}$.
$^{g}$ The reflection fraction; see Section 3 and references therein. 
$^{h}$ The component normalization; see Section 3 and references therein.
$^{j}$ The degrees of freedom when fitting the given model as shown here and described in Section 3.  There were 1248 energy bins in total for every fit.
$^{\dag}$ These values were fixed, as ``pexriv'' was not able to constrain the ionization.  Our favored model is the constant density ionized disc model (CDID), which was able to simultaneously fit the reflection fraction and ionization.}
\end{table}

\begin{figure*}
\centerline{
\includegraphics[width=0.80\textwidth]{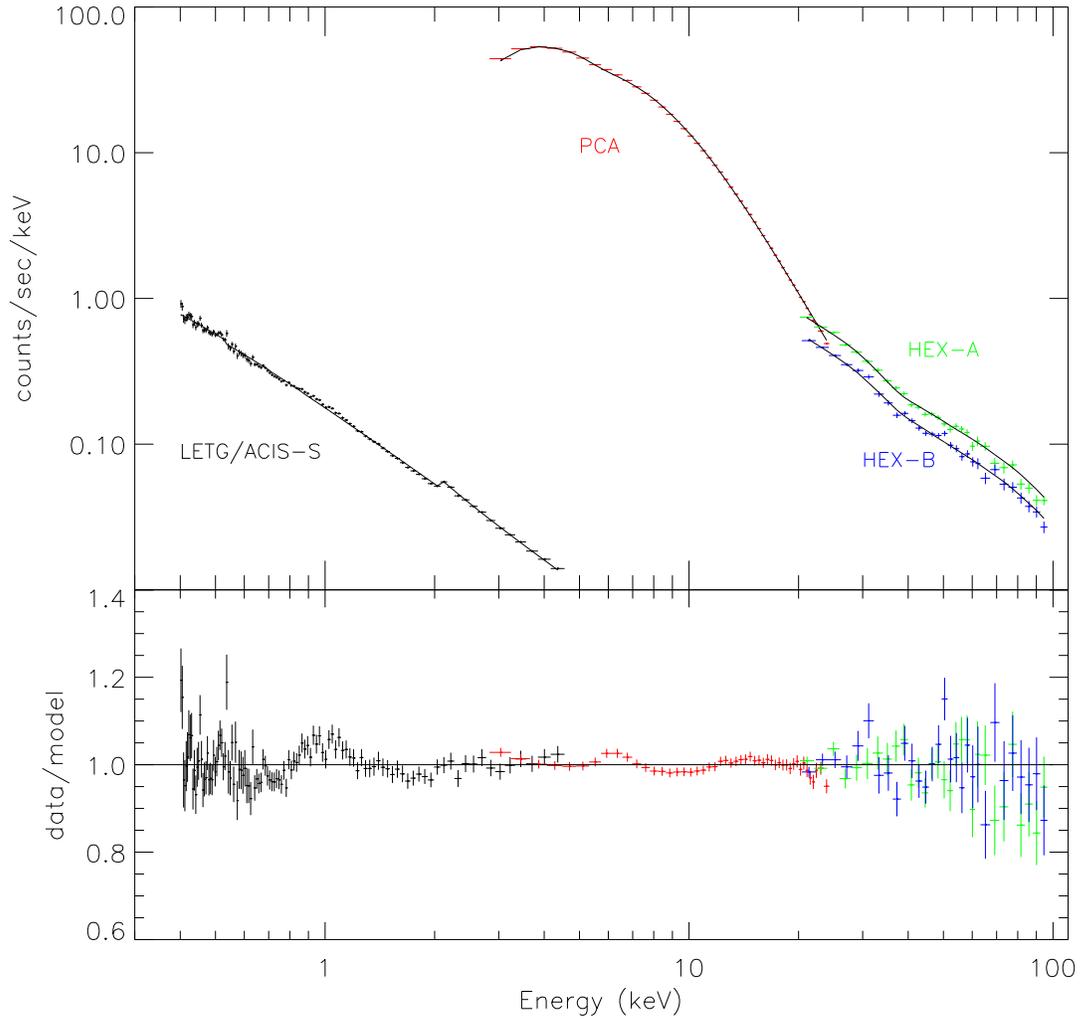}
}
\caption{Spectra from XTE~J1118$+$480 fit simultaneously with a simple
power-law model, and the data/model ratio for that fit.  Due to the
small reflection fraction, fits with the reflection models are
indistinguishable from that shown here.  Pictured above: in black, the
\textit{Chandra} LETG/ACIS-S spectrum (rebinned by a factor of 10 for
clarity); in red, the \textit{RXTE} PCA spectrum; and in green and
blue (respectively), the \textit{RXTE} HEXTE-A and HEXTE-B spectra.}
\end{figure*}

\begin{figure*}
\centerline{
\includegraphics[width=0.80\textwidth]{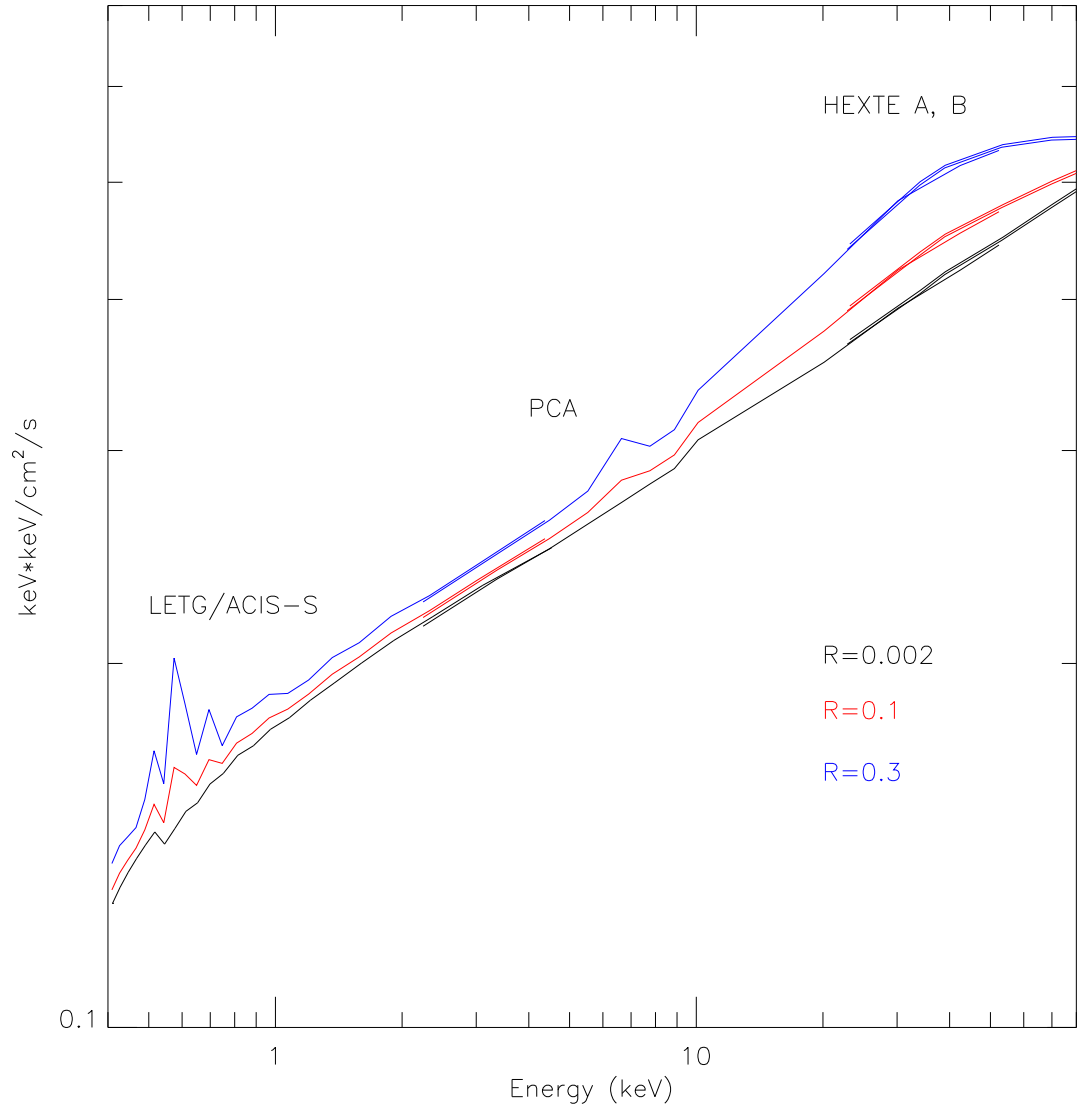}
}
\caption{Different reflection fractions produce different curvatures
($R$ is the fraction of the observed flux that was reflected).  In
fits with the constant-density ionized disc model, a reflection
fraction of $R=0$ is preferred.  Here, we have plotted the 90 per cent
confidence upper-limit $R=0.002$.  Plotted in red and blue (assuming
the same unreflected normalization): the model for $R=0.1$ and
$R=0.3$, respectively.  Models are rebinned by a factor of 50 for
visual clarity.  Note that within the text, we discuss why an upper
limit of $R\leq0.005$ may be more appropriate.}
\end{figure*}


\begin{thebibliography}{}

\bibitem[\protect\citeauthoryear{Arnaud}{1996}]{arn96}Arnaud K.A.,
1996, in Jacoby G., Barnes J., eds, Astronomical Data Analysis
Software and Systems V, ASP Conference Series, 101, 17

\bibitem[\protect\citeauthoryear{Bel}{1999}]{b99}Beloborodov, A. M.,
1999, ApJ, 510L, 123B

\bibitem[\protect\citeauthoryear{Esin}{1997}]{esn97}Esin, A. A.,
McClintock, J. E., and Narayan, R., 1997, ApJ, 489, 865

\bibitem[\protect\citeauthoryear{Esin}{1998}]{esn98}Esin, A. A.,
Narayan, R., Cui, W., Grove, E. J., and Zhang, S., 1998, ApJ 505, 854

\bibitem[\protect\citeauthoryear{Esin}{2001}]{esn01}Esin, A. A.,
McClintock, J. E., Drake, J. J., Garcia, M. R., Haswell, C. A., Hynes,
R. I., and Muno, M. P., 2001, ApJ, 555, 483

\bibitem[\protect\citeauthoryear{Fender}{2001}]{f01}Fender, R. P.,
Hjellming, R. M, Tilanus, R. J., et al., 2001, MNRAS, 322, L32

\bibitem[\protect\citeauthoryear{Gierlinski}{1997}]{gl97}Gierlinski,
M., et al., 2001, MNRAS 288, 958

\bibitem[\protect\citeauthoryear{Gierlinski}{1999}]{gl99}Gierlinski,
M., et al.\ 2001, MNRAS 288, 958


\bibitem[\protect\citeauthoryear{George}{1991}]{g91}George, I. M., \&
Fabian, A. C., 1991, MNRAS, 249, 352

\bibitem[\protect\citeauthoryear{Fro}{2001}]{fro01}Frontera, F., et
al., 2001, ApJ, in press, astro-ph/0107199 

\bibitem[\protect\citeauthoryear{Gie}{1997}]{gie97}Gierlinski, M., et
al., 1997, MNRAS, 288, 958

\bibitem[\protect\citeauthoryear{Gilfanov}{2000}]{gil00}Gilfanov, M.,
Churazov, E., and Revnivtsev, M., 2000, Proc. of the 5th CAS/MPG
Workshop on High Energy AStrophysics, astro-ph/0002415

\bibitem[\protect\citeauthoryear{Jahoda}{2000}]{jah00}Jahoda, K.,
2000, unpublished talk presented 2000 March 22 at the Rossi 2000
Symp. (Greenbelt, MD:GSFC/NASA)

\bibitem[\protect\citeauthoryear{Juett}{2002}]{jue02}Juett, A.,
et al., 2002, ApJL, subm.

\bibitem[\protect\citeauthoryear{Mag}{1995}]{mag95}Magdziarz, P., and
Zdziarski, A. A., 1995, MNRAS, 273, 837

\bibitem[\protect\citeauthoryear{Markoff}{2001}]{mar01}Markoff, S.,
Falcke, H., and Fender, R., 2001, A \& A, 272L, 25

\bibitem[\protect\citeauthoryear{McClintock}{2001a}]{mc01a}McClintock,
J. E., et al., 2001a, ApJ, 555, 477

\bibitem[\protect\citeauthoryear{McClintock}{2001b}]{mc01b}McClintock,
J. E., Garcia, M. R., Caldwell, N., Falco, E. E., Garnavich, P. M.,
and Zhao, P., 2001, ApJ, 551, L147

\bibitem[\protect\citeauthoryear{Mitsuda}{1984}]{mit84}Mitsuda, K., et
al., 1984, PASJ, 36, 741

\bibitem[\protect\citeauthoryear{Merloni}{2000}]{mer01}Merloni, A.,
Di~Matteo, T., and Fabian, A. C., 2000, MNRAS, 318, L15

\bibitem[\protect\citeauthoryear{Miller}{2001}]{mil01}Miller, J. M.,
et al., 2001, ApJ, 546, 1055

\bibitem[\protect\citeauthoryear{Miller}{2002}]{mil02}Miller, J. M.,
et al., 2002, ApJ, subm., astro-ph/0202083

\bibitem[\protect\citeauthoryear{Narayan}{1994}]{nar94}Narayan, R.,
and Yi, I., 1994, ApJ, 428L, 13

\bibitem[\protect\citeauthoryear{Nay}{2001}]{nay01}Nayakshin, S.,
Kallman, T. R., and Kazanas, D., 2001, available at
http:\/\/lheawww.gsfc.nasa.gov\/users\/serg\/ms.ps

\bibitem[\protect\citeauthoryear{Patel}{2001}]{pat01}Patel, S. K.,
et al., 2001, ApJL, in press, astro-ph/0110182

\bibitem[\protect\citeauthoryear{Petrucci}{2001}]{pet01}Petrucci, P. O.,
Merloni, A., Fabian, A., Haardt, F., and Gallo, E., 2001, MNRAS subm.,
astro-ph/0108342

\bibitem[\protect\citeauthoryear{Ross}{1993}]{r93}Ross, R. R., \&
Fabian, A. C., 1993, MNRAS, 261, 74

\bibitem[\protect\citeauthoryear{Ross}{1999}]{rfy99}Ross, R. R.,
Fabian, A. C., \& Young, A. J., 1999, MNRAS, 306, 461

\bibitem[\protect\citeauthoryear{Wagner}{2001}]{wag01}Wagner, R. M.,
Foltz, C. S., Shahbaz, T., Casares, J., Charles, P. A., Starrfield,
S. G., and Hewett, P., 2001, ApJ, 556, 42

\bibitem[\protect\citeauthoryear{Zdz}{1998}]{zdz98}Zdziarski, A. A.,
Poutanen, J., Mikolajewska, J., Gierlinski, M., Ebisawa, K., and
Johnson, W. N., 1998, MNRAS, 301, 435

\bibitem[\protect\citeauthoryear{Zycki}{1997}]{zy00}Zycki, P. T.,
Done, C., and Smith, D. A., 1997, ApJ, 488, L113


\end{thebibliography}
\end{document}